\documentclass[fleqn]{article}
\usepackage{amsmath,amssymb}
\usepackage{epsfig,graphicx}

\renewcommand{\theequation}{\arabic{section}.\arabic{equation}}
\renewcommand{\thesection}{\arabic{section}.}

\catcode`@=11 \@addtoreset{equation}{section} \catcode`@=12

\begin{document}
\title{\vskip-1.7cm \bf  Thermal power spectrum in the CFT driven cosmology}
\date{}
\author{A.O.Barvinsky}
\maketitle
\hspace{-8mm} {\,\,\em Theory Department, Lebedev
Physics Institute, Leninsky Prospect 53, Moscow 119991, Russia}

\begin{abstract}
We present an overview of the recently suggested cosmological model driven by conformal field theory (CFT) with the initial  conditions in the form of the microcanonical density matrix. In particular, we discuss the origin of inflationary stage in this model and a novel feature -- the thermal nature of the primordial power spectrum of the CMB anisotropy. The relevant effect of ``temperature of the relic temperature anisotropy'' can be responsible for a thermal contribution to the red tilt of this spectrum, additional to its conventional vacuum component. The amplification of this effect due to recently established $a$-theorem in CFT is briefly discussed.
\end{abstract}

\maketitle
\section{Introduction}
The purpose of this paper is to consider certain aspects of the CMB spectrum in the recently suggested model of CFT driven cosmology \cite{slih,why}. This model represents the synthesis of two main ideas in the theory of the early Universe -- new concept of the cosmological microcanonical density matrix as the initial state of the Universe and application of this concept to the system with a large number of quantum fields conformally coupled to gravity. It plays important role within the cosmological constant and dark energy problems. In particular, its statistical ensemble is bounded to a finite range of values of the effective cosmological constant, it incorporates inflationary stage and is potentially capable of generating the cosmological acceleration phenomenon within the so-called Big Boost scenario \cite{slih,bigboost}. Moreover, as was noticed in \cite{DGP/CFT}, the CFT driven cosmology provides perhaps the first example of the initial quantum state of the inflationary Universe, which has a thermal nature of the primordial power spectrum of cosmological perturbations. This suggests a new mechanism for the red tilt of the CMB anisotropy, complementary to the conventional mechanism which is based on a small deviation of the inflationary expansion from the exact de Sitter evolution \cite{ChibisovMukhanov}.

Thus, the CFT driven cosmology revives a certain version of hot Big Bang origin of the Universe. Whereas the inflation paradigm eliminated Big Bang by replacing its singular state by the initial vacuum whose quantum fluctuations eventually generate the present large scale structure, this model again incorporates an effectively thermal state at the onset of the cosmological evolution. The rationale behind this is the fact that selection of the concrete pure vacuum state is subject to a great ambiguity, whereas the microcanonical ensemble sums over all possible selections. Creation of the Universe from {\em everything} \cite{why} is conceptually more appealing than creation from {\em nothing} \cite{noboundary,tunnel}, because the democracy of the microcanonical equipartition better fits the principle of the Occam razor than the selection of a concrete state.

It is well known that universality of the inflationary CMB spectrum follows from the short-wavelengths part of the cosmological perturbations whose vacuum state should be of the Hadamard type \cite{BunchDavis,ChibisovMukhanov}. However, experimental refinement of the CMB observations \cite{WMAP9} leads to growing interest in modified vacuum and non-vacuum states of cosmological perturbations, because these states might \cite{{nongaussianity},GancKomatsu} or might not \cite{Ashoorioonetal} generate enhanced non-gaussianities of the CMB bispectrum. This in turn suggests the search for models in which this choice is not based on some ad hoc assumptions, but rather follows from fundamental first principles. The model of CFT driven cosmology of \cite{slih,why} is perhaps the first example of such a cosmological setup.

This setup has a clear origin in terms of operator quantization of gravity theory in the Lorentzian signature spacetime and is based on a natural notion of the microcanonical density matrix as a projector on the space of solutions of the quantum gravitational Dirac constraints -- the system of the Wheeler-DeWitt equations \cite{why,whyBFV}. Its statistical sum has a representation of the Euclidean quantum gravity (EQG) path integral \cite{slih,why}
    \begin{eqnarray}
    &&Z=
    \!\!\int\limits_{\,\,\rm periodic}
    \!\!\!\! D[\,g_{\mu\nu},\phi\,]\;
    e^{-S[\,g_{\mu\nu},\phi\,]},         \label{Z}
    \end{eqnarray}
over metric $g_{\mu\nu}$ and matter fields $\phi$ which are
periodic on the Euclidean spacetime with a time compactified to a circle $S^1$.

As shown in \cite{slih,why}, this statistical sum is approximately calculable and has a good predictive power in the gravitational model with the primordial cosmological constant $\varLambda$ and the matter sector which mainly consists of a large number $\mathbb{N}$ of free (linear) fields $\phi$ conformally coupled to gravity -- conformal field theory (CFT) with the action $S_{CFT}[\,g_{\mu\nu},\phi\,]$,
    \begin{eqnarray}
    &&S[\,g_{\mu\nu},\phi\,]=-\frac1{16\pi G}
    \int d^4x\,g^{1/2}\,(R-2\varLambda)
    +S_{CFT}[\,g_{\mu\nu},\phi\,].    \label{tree}
    \end{eqnarray}
Important point, which allows one to overstep the limits of the usual semiclassical expansion, consists here in the possibility to omit the integration over conformally non-invariant matter fields and spatially-inhomogeneous metric modes on top of a dominant contribution of numerous conformal species. Integrating them out one obtains the effective gravitational action $S_{\rm eff}[\,g_{\mu\nu}]$,
    \begin{eqnarray}
    &&S_{\rm eff}[\,g_{\mu\nu}]=-\frac1{16\pi G}
    \int d^4x\,g^{1/2}\,(R-2\varLambda)
    +\varGamma[\,g_{\mu\nu}],    \label{effective}\\
    &&e^{-\varGamma[\,g_{\mu\nu}]}=\int D\phi\,e^{-S_{CFT}[\,g_{\mu\nu},\phi\,]}.
    \end{eqnarray}
When applied to a spatially closed model with the topology of a 3-dimensional sphere, the actual calculation of $Z$ is based on disentangling the minisuperspace sector of the homogeneous Friedmann-Robertson-Walker (FRW) metric, $g_{\mu\nu}(x)\to [\,a(\tau),N(\tau);\,h_{\mu\nu}(x)\,]$,
    \begin{eqnarray}
    g_{\mu\nu}^{FRW} dx^\mu dx^\nu=N^2(\tau)\,d\tau^2
    +a^2(\tau)\,d^2\Omega^{(3)},              \label{FRW}
    \end{eqnarray}
and the transition in (\ref{Z}) to the form of the path integral over a minisuperspace lapse function $N(\tau)$ and scale factor $a(\tau)$ of this metric,
    \begin{eqnarray}
    &&Z=
    \!\!\int\limits_{\,\,\rm periodic}
    \!\!\!\! Dg_{\mu\nu}\;
    e^{-S_{\rm eff}[\,g_{\mu\nu}]}\simeq
    \!\!\int\limits_{\,\,\rm periodic}
    \!\!\!\!
     D[\,a,N\,]\;
    e^{-S_{\rm eff}[\,a,\,N\,]}.  \label{miniZ}
    \end{eqnarray}
Here $S_{\rm eff}[\,a,\,N\,]\equiv S_{\rm eff}[\,g_{\mu\nu}^{FRW}]$ is the effective action calculated on the FRW background, $D[\,a,N\,]$ is assumed to include the Faddeev-Popov gauge fixing for the one-dimensional diffeomorphism invariance of the metric (\ref{FRW}) \cite{PIQC} and the integration over the graviton sector of $h_{\mu\nu}(x)$, similarly to other non-conformal modes, is discarded in the leading order of the relevant $1/\mathbb{N}$ expansion.\footnote{Note that the integration over minisuperspace variables $N(\tau)$ and $a(\tau)$ should be retained. They comprise a modular rather than a local degree of freedom and provide the microcanonical nature of the problem. Omission of integration over $N(\tau)$ and $a(\tau)$ would result in a kinematically incorrect and incomplete definition of the statistical sum. In fact, semiclassically this integration results in the (quasi)thermal nature of the ensemble with a definite temperature parameter, because its inverse -- the period of the Euclidean time of saddle-point configurations for (\ref{miniZ}) -- follows from this integration (see below).}

The power of the minisuperspace integral (\ref{miniZ}) is that its effective action, $S_{\rm eff}[\,a,\,N\,]$, is exactly calculable by the conformal transformation converting (\ref{FRW}) into the static Einstein metric with $a={\rm const}$. It becomes the sum of the contribution of this conformal transformation \cite{FHHS,anomalyaction}, determined by the well-known conformal anomaly of a quantum CFT in the external gravitational field \cite{Duffanomaly}
    \begin{eqnarray}
    &&g_{\mu\nu}\frac{\delta
    \varGamma}{\delta g_{\mu\nu}} =
    \frac{1}{4(4\pi)^2}g^{1/2}
    \left(\alpha \Box R +
    \beta E +
    \gamma C_{\mu\nu\alpha\beta}^2\right), \label{anomaly}\\
    &&E=R_{\mu\nu\alpha\gamma}^2-4R_{\mu\nu}^2
    + R^2,                                \label{GaussBonnet}
    \end{eqnarray}
and the contribution of a static Einstein Universe. The latter is the combination of the vacuum (Casimir) energy part \cite{Casimir} and free energy of a typical boson or fermion statistical sum \cite{slih}.

Physics of the CFT driven cosmology is entirely determined by this effective action. Solutions of its equations of motion, which give a dominant contribution to the statistical sum, turn out to be the so-called {\em garlands} -- the cosmological instantons of $S^1\times S^3$ topology, which have the periodic scale factor $a(\tau)$ -- the function of the Euclidean time belonging to the circle $S^1$ \cite{slih}. These instantons serve as initial conditions for the cosmological evolution $a_L(t)$ in the physical Lorentzian spacetime. The latter follows from $a(\tau)$ by the analytic continuation $a_L(t)=a(\tau_++it)$ to the complex plane of the Euclidean time at the point of the maximum value of the Euclidean scale factor $a_+=a(\tau_+)$. As it was discussed in \cite{bigboost,DGP/CFT} this Lorentzian evolution can incorporate a finite inflationary stage if the model (\ref{tree}) is generalized to the case when a primordial cosmological constant $\varLambda$ is replaced by a composite operator $\varLambda(\phi)$. This is the potential of the inflaton field $\phi$ staying in the slow-roll regime during the Euclidean and inflationary stages and decaying in the end of inflation by a usual exit scenario.

All this is recapitulated in Sects. 2 and 3 below before we pass to the discussion of the primordial CMB fluctuations generated during this inflationary stage. This power spectrum has two important distinctions from conventional inflationary models. First, it is dynamically more complicated because the gravitational sector of the model even in the scalar sector of cosmological perturbations is essentially different from Einstein theory. Second, this spectrum has a thermal (or quasi-thermal) nature in view of the thermal rather than vacuum nature of initial conditions. A compact range of the Euclidean time $\tau$ results in the thermal contribution to the spectrum with the temperature defined by the time period of the instanton configuration.\footnote{This contribution should not be confused with the thermal nature of the CMB observations of $\Delta T/T$, measuring the temperature of the relic radiation. In our model $\Delta T/T$ becomes a subject of thermal distribution, so that in fact we have a {\em temperature of the CMB temperature}.} In particular, it enhances the red tilt of the primordial spectrum, the vacuum part of which is caused by the slow-roll deviation of the inflationary evolution from the exact de Sitter expansion \cite{ChibisovMukhanov}. So in Sect. 3 we will focus on this thermal part of the CMB spectrum.

Unfortunately, it turns out to be exponentially red shifted by
a large curvature to horizon scale ratio $H_0a_0=(\Omega_0-1)^{-1/2}$ and observationally negligible for all low-spin CFT models. However, the thermal part also depends on an average specific value per one conformal degree of freedom of the parameter $\beta$ -- the coefficient of the Gauss-Bonnet term in the anomaly (\ref{anomaly}) -- and grows with the growing $\beta$. Thus, for hypothetical higher-spin conformal fields it can take arbitrarily high values \cite{ChristensenDuff,Tseytlinconf} and provide a considerable thermal effect .

A possible mechanism of growing $\beta$ can be attributed to the recently discovered $a$-theorem for renormalization group (RG) flow in interacting field theory with broken conformal invariance \cite{KomargodskiSchwimmer,Komargodski}. For free lowest-order conformal spins, $s=0,1/2,1$, $\beta$ is an additive sum of their one-loop contributions
    \begin{eqnarray}
    \beta=\frac1{180}\,\big(\mathbb{N}_0+11 \mathbb{N}_{1/2}+
    62 \mathbb{N}_{1}\big),                \label{100}
    \end{eqnarray}
where $\mathbb{N}_s$ are the numbers of spin-$s$ species ($\mathbb{N}_{1/2}$ is the number of Dirac spinors). Remarkable property of $\beta$ is that it becomes a running parameter $\beta=\beta(\mu)$ monotonically growing from the infrared to ultraviolet regime along the RG flow, $\beta(\infty)-\beta(0)>0$. This comprises the so-called $a$-theorem \cite{KomargodskiSchwimmer,Komargodski} stemming from the anomaly matching concept in \cite{matching}.

The virtue of this theorem is that the cosmological expansion can be associated with the RG flow from deep UV dominated by conformal higher spin fields to the current low spin IR regime. Moderately small present value of $\beta$ can be interpreted as a result of such evolution from a larger initial value $\beta_{UV}$ responsible for the formation of a considerable thermal part of CMB. We discuss this mechanism in Sect.4, where in particular we dwell on the status of the dilaton mode and dilaton effective action in CFT cosmology vs that of \cite{KomargodskiSchwimmer,Komargodski}. The main difference is that, whereas in \cite{KomargodskiSchwimmer,Komargodski} the metric and dilaton variables are just auxiliary fields probing quantum CFT models, in cosmology these fields comprise important physical observables. We finish Sect.4 by discussing the non-dynamical status of the scalar sector of these variables (similar to the Einstein theory with the conformal mode dynamically eliminated by the Hamiltonian constraint), their causal structure induced by the dilaton to gravity braiding and peculiarities of the UV renormalization in the CFT driven cosmology which underlies certain universality properties of $\beta$-dependence in this model of the early quantum Universe. Final Sect. 5 summarizes our conclusions and future prospects.

\section{Thermodynamics of the initial state in the CFT cosmology}

For closed cosmology with $S^3$ topology of its spatial section its minisuperspace effective action $S_{\rm eff}[\,g_{\mu\nu}^{FRW}]\equiv S_{\rm eff}[\,a,N\,]$ reads in units of the Planck mass $m_P=(3\pi/4G)^{1/2}$ \cite{slih}
    \begin{eqnarray}
    &&S_{\rm eff}[\,a,N\,]=m_P^2\int_{S^1} d\tau\,N \left\{-aa'^2
    -a+ \frac\varLambda3 a^3+\,B\left(\frac{a'^2}{a}
    -\frac{a'^4}{6 a}\right)+\frac{B}{2a}\,\right\}
    +F(\eta),                              \label{effaction}\\
    &&F(\eta)=\pm\sum_{\omega}\ln\big(1\mp
    e^{-\omega\eta}\big),                 \label{freeenergy}\\
    &&\eta=\int_{S^1} \frac{d\tau N}a,     \label{period}
    \end{eqnarray}
where $a'\equiv da/Nd\tau$. The first three terms in curly brackets of (\ref{effaction}) represent the Einstein action with a primordial (but renormalized by quantum corrections) cosmological constant $\varLambda\equiv 3H^2$ ($H$ is the corresponding Hubble constant). The terms proportional to the constant $B$ correspond to the contribution of the conformal anomaly and the contribution of the vacuum (Casimir) energy $(B/2a)$ on a conformally related static Einstein spacetime, discussed in Introduction. Finally, $F(\eta)$ is the free energy of conformal fields also coming from this Einstein space -- a typical boson or fermion sum over CFT field oscillators with energies $\omega$ on a unit 3-sphere, $\eta$ playing the role of the inverse temperature --- an overall circumference of the $S^1\times S^3$ instanton in terms of the conformal time (\ref{period}).

The constant $B$,
    \begin{eqnarray}
    B=\frac{3\beta}{4 m_P^2},         \label{B}
    \end{eqnarray}
is determined by the coefficient $\beta$ of the topological Gauss-Bonnet invariant $E$ in the overall conformal anomaly
(\ref{anomaly}). For free low-spin fields it is always positive \cite{Duffanomaly} and is defined by the one-loop expression (\ref{100}). The UV ambiguous coefficient $\alpha$ is renormalized  to zero by a local counterterm $\sim \alpha R^2$ which guarantees the absence of higher derivative terms in the action (\ref{effaction}). This automatically gives the renormalized Casimir energy the value $m_P^2B/2a=3\beta/8a$ which universally expresses in terms of the same coefficient in the conformal anomaly \cite{universality}.\footnote{This
universality property follows from the fact that in a static Einstein Universe of the size $a$ the Casimir energy of conformal fields is determined by the conformal anomaly coefficients and equals $(3\beta-\alpha/2)/8a$ \cite{universality} (see discussion in Sect.4.)} The coefficient $\gamma$ of the Weyl tensor squared term $C^2_{\mu\nu\alpha\beta}$ does not enter the expression (\ref{effaction}) because $C_{\mu\nu\alpha\beta}$ identically vanishes for any FRW metric.

Semiclassically the statistical sum (\ref{miniZ}) is dominated by the solutions of the effective equation for the action (\ref{effaction}), $\delta S_{\rm eff}/\delta N(\tau)=0$. This is the modification of the Euclidean Friedmann equation,
    \begin{eqnarray}
    &&-\frac{a'^2}{a^2}+\frac{1}{a^2}
    -B \left(\,\frac{a'^4}{2a^4}
    -\frac{a'^2}{a^4}\right) =
    \frac\varLambda3+\frac{C}{ a^4},     \label{efeq}\\
    &&m_P^2 C =
    \frac{m_P^2 B}2+\frac{dF}{d\eta},  \label{bootstrap}
    \end{eqnarray}
by the anomalous $B$-term and the radiation term $C/a^4$. The constant $C$ here characterizes the sum of the Casimir energy $(\sim B/2)$ and the energy of the gas of thermally excited particles with the inverse temperature $\eta$ given by (\ref{period}),
    \begin{eqnarray}
    \frac{dF}{d\eta}=\sum_\omega\frac{\omega}
    {e^{\omega\eta}\mp 1}.
    \end{eqnarray}

As shown in \cite{slih,why,DGP/CFT} the solutions of this integro-differential equation\footnote{Note that the constant $C$ is a nonlocal functional of the history $a(\tau)$ -- Eq.(\ref{bootstrap}) plays the role of the bootstrap equation for the amount of radiation determined by the background on top of which this radiation evolves and produces back reaction.} give rise to the set of periodic $S^3\times S^1$ instantons with the oscillating scale factor -- {\em garlands} that can be regarded as the thermal version of the Hartle-Hawking instantons. The scale factor oscillates $\kappa$ times ($\kappa=1,2,3,...$) between the maximum and minimum values
$a_\pm=a(\tau_\pm)$, $a_-\leq a(\tau)\leq a_+$,
    \begin{eqnarray}
    a^2_\pm=\frac{1\pm\sqrt{1-4H^2C}}{2H^2},     \label{apm}
    \end{eqnarray}
so that the full period of the conformal time (\ref{period}) is the $2\kappa$-multiple of the integral between the two neighboring turning points of $a(\tau)$, $\dot a(\tau_\pm)=0$,
    \begin{eqnarray}
    &&\eta=2\kappa\int_{\tau_-}^{\tau_+}
    \frac{d\tau\,N}{a}=2\kappa\int_{a_-}^{a_+}
    \frac{da}{a'a}.                       \label{period1}
    \end{eqnarray}
This value of $\eta$ is finite and determines a finite effective temperature $T=1/\eta$ as a function of $G=3\pi/4m_P^2$ and $\varLambda=3H^2$. This is the artifact of a microcanonical ensemble in cosmology \cite{why} with only two freely specifiable dimensional parameters --- the renormalized gravitational and renormalized cosmological constants.

These $S^3\times S^1$ garland-type instantons exist only in the limited range of the cosmological constant $\varLambda=3H^2$ \cite{slih} and belong to the curvilinear domain in the two-dimensional plane of the Hubble constant $H^2$ and the amount of radiation constant $C$ (each instanton being represented by a point in this plane),
    \begin{eqnarray}
    &&0<\varLambda_{\rm min}<\varLambda<
    \varLambda_{\rm max}=\frac3{2B},     \label{landscape}\\
    &&B-B^2H^2\leq C\leq \frac1{4H^2}.   \label{domain}
    \end{eqnarray}
In this domain they form an countable, $\kappa=0,1,2,...$, sequence of one-parameter families -- curves interpolating
between the lower straight line
boundary $C=B-B^2H^2$ and the upper hyperbolic boundary $C=1/4H^2$. Each curve corresponds to a respective $\kappa$-fold instantons of the above type. Therefore, the spectrum of admissible values of $\varLambda$ has a band structure, each band being a projection of the $\kappa$-curve to the $H^2$ axis. The sequence of bands of ever narrowing widths with $\kappa\to\infty$ accumulates at the upper bound of this range $H^2_{\rm max}=1/2B$. The lower bound $H^2_{\rm min}$ -- the lowest point of $\kappa=1$ family -- can be obtained numerically for any field content of the model.

For a large number of conformal fields $\mathbb{N}$, and therefore a large $\beta\propto \mathbb{N}$, the both bounds are of the order $H^2_{\rm min}\sim m_P^2/\mathbb{N}$. Thus the restriction (\ref{landscape}) suggests a kind of $1/\mathbb{N}$ solution of the cosmological constant problem, because specifying a sufficiently high number of conformal fields one can achieve a primordial value of $\varLambda$ well below the Planck scale where the effective theory applies, but high enough to generate a sufficiently long inflationary stage.

Important property is that at the upper boundary of the domain (\ref{domain}), the instantons are static with $a(\tau)=a_+=a_-$. They represent thermodynamical equilibrium with the temperature inverse proportional to the period (\ref{period1}), $T=1/a\eta$, which is exactly calculable and reads \cite{slih}
    \begin{eqnarray}
    \eta=\pi \kappa \sqrt{2(1-2BH^2)}.         \label{upperb}
    \end{eqnarray}
Then from the bootstrap equation (\ref{bootstrap}) and $C=1/4H^2$ it follows that the Hubble factor and conformal time period of these instantons are given by
    \begin{eqnarray}
    &&H^2=\frac{m_P^2}{2Bm_P^2
    +4\,\frac{dF}{d\eta}},       \label{Hcool}\\
    &&\eta=2\pi \kappa\sqrt{\frac{\frac1{Bm_P^2}
    \frac{dF}{d\eta}}{1+
    \frac2{Bm_P^2}\frac{dF}{d\eta}}}. \label{etaequation}
    \end{eqnarray}
The last expression serves as the equation for $\eta$ and immediately yields the bound $\eta<\pi \kappa\sqrt2$.

Other instantons in the domain (\ref{domain}) slightly violate thermodynamic equilibrium, but do not qualitatively change the situation. Therefore we will basically consider the exactly solvable case with (\ref{upperb})-(\ref{etaequation}). In what follows we present two analytically tractable limits of low and high temperatures.

\subsection{Low temperature limit}
Low temperature limit corresponds to large $\eta$ when the free energy $F(\eta)$ is dominated by the first term of the statistical sum (\ref{freeenergy})
    \begin{eqnarray}
    F\simeq -\mathbb{N}_0 d_0 e^{-\omega_0\eta},
    \quad\frac{dF}{d\eta}\simeq \mathbb{N}_0
    \omega_0 d_0 e^{-\omega_0\eta},
    \quad \omega_0=1,\quad d_0=1.     \label{truncation}
    \end{eqnarray}
Here the subscript zero denotes the contribution of $\mathbb{N}_0$ scalar conformal degrees of freedom whose energy of the lowest field-theoretical oscillator on the 3-dimensional sphere of the unit radius $\omega_0=1$ and its degeneracy $d_0=1$. All other fields have higher values of lowest energy and therefore add to (\ref{truncation}) exponentially smaller contributions unless their multiplicities $\mathbb{N}_s$ outnumber their small Boltzmann weights.\footnote{In models without conformal scalars Eq.(\ref{truncation}) is obviously replaced by the one with the lowest value of $\omega_0>1$ and $d_0>1$ of its higher spin fields.} Therefore, the ratio of the thermal radiation energy to the vacuum energy equals
    \begin{eqnarray}
    \frac2{Bm_P^2}\frac{dF}{d\eta}=\frac{8}{3\bar\beta} e^{-\eta},
    \end{eqnarray}
where $\bar\beta$ is a specific value of the coefficient $\beta$ per one {\em scalar} conformal degree of freedom
    \begin{eqnarray}
    \bar\beta=\frac\beta{\mathbb{N}_0 }
    \end{eqnarray}
Thus, Eq.(\ref{etaequation}) for $\eta$ takes the form
    \begin{eqnarray}
    e^{\eta}=\frac{16}{3\bar\beta}
    \left(\frac{\kappa^2\pi^2}{\eta^2}-\frac12\right)
    \end{eqnarray}
which for $\kappa\gg 1$ has the asymptotic solution
    \begin{eqnarray}
    \eta\simeq\ln\left(\frac{16\pi^2\kappa^2}{3\bar\beta}
    \right).                    \label{cooleta}
    \end{eqnarray}
The cosmological constants of the corresponding instantons $\Lambda=3H^2$ in view of (\ref{Hcool}) are very close to the maximal value -- quantum gravity scale of the model
    \begin{eqnarray}
    H^2=\frac1{2B}\left(1+\frac1{2\pi^2 \kappa^2}\right)^{-1}.
    \end{eqnarray}
Their physical temperature $T=1/a\eta$ in view of $a=a_\pm=\sqrt{B}$ decreases with the growing $\bar\beta$ (in contrast to the ``comoving" temperature $1/\eta$)
    \begin{eqnarray}
    T=\frac{2 m_P}{\sqrt{3\bar\beta\,\mathbb{N}_0}\,
    \ln\big(16\pi^2\kappa^2/3\bar\beta\big)}.     \label{Tlow}
    \end{eqnarray}
This will be the temperature of thermal corrections to the CMB spectrum, but as we will see due to red shifting the observable corrections to this spectrum in the long wavelengths part will actually be determined by the comoving temperature.

For all lowest order spins $s\leq 1$ belonging to the range $1\leq 180\beta_s\leq 62$ this is a good approximation, because the truncation of thermal sums (\ref{truncation}) is justified by the bound $e^{-\eta}\simeq 3\beta/16\pi^2\kappa^2\ll 1$. Interestingly, it remains good even for the single-fold instanton with $\kappa=1$.

\subsection{High temperature limit}
In the system with $\mathbb{N}_s$ conformal multiplets of low spins $s=0,1/2,1$ the high temperature behavior of their total free energy, $\eta\ll 1$, reads
    \begin{eqnarray}
    &&F\simeq -\mathbb{N}\frac{\pi^4}{45\eta^3}, \quad
    \frac{dF}{d\eta}\simeq \mathbb{N}
    \frac{\pi^4}{15\eta^4},       \label{truncationhot}\\
    &&\mathbb{N}=\mathbb{N}_0
    +2\left(\frac78\,\mathbb{N}_{1/2}+\mathbb{N}_1\right),
    \end{eqnarray}
where $\mathbb{N}$ is the effective number of their degrees of freedom with two polarizations per Majorana spinor and vector multiplets (modified by the well-known coefficient 7/8 distinguishing the thermal contribution of fermions vs that of bosons). Therefore, the ratio of the thermal radiation energy to the vacuum energy equals
    \begin{eqnarray}
    \frac2{Bm_P^2}\frac{dF}{d\eta}=\frac{8\pi^4}{45\eta^4 \tilde\beta},
    \end{eqnarray}
where $\tilde\beta$ is the average (specific per one conformal degree of freedom) value of the parameter $\beta_s$
    \begin{eqnarray}
    \tilde\beta=\frac1{\mathbb{N}}\,\sum_s \beta_s \mathbb{N}_s.
    \end{eqnarray}

Then for the variable $x=(\eta/2\pi\kappa)^2$ Eq.(\ref{etaequation}) immediately leads to the equation
    \begin{eqnarray}
    x^3+\frac1\lambda\, x=\frac1{2\lambda}
    \end{eqnarray}
where $\lambda\equiv 90\, \kappa^4 \tilde\beta$. For large $\lambda\gg 1$ its solution $x\simeq(1/2\lambda)^{1/3}$ gives
    \begin{eqnarray}
    &&\eta=2\pi\left(\frac{\kappa^2}
    {180 \tilde\beta}\right)^{1/6},      \label{etahot}\\
    &&H^2=\frac1{2B}\,
    \left(1+\frac2{(180\,\kappa^4 \tilde\beta)^{1/3}}\right)^{-1},\\
    &&T=\frac{m_P}{\pi\sqrt\mathbb{N}}\,
    \left(\frac{20}{3\tilde\beta^2\kappa^2}\right)^{1/6}
    \end{eqnarray}

This approximation works well only in the limit $\eta\ll 1$, which implies $\kappa\ll (180\,\tilde\beta)^{1/2}/(2\pi)^3$ and because of $\kappa\geq 1$ makes sense only when $180\, \tilde\beta\gg 64\pi^6$. Thus, the high temperature limit corresponds to large $\beta$ and relatively small $\kappa$. The ``comoving" temperature $1/\eta$ grows with $\beta$ and decreases with $\kappa$, as it should because multi-fold instantons are always ``cooler" than the single fold one. Note that again the physical temperature $T=1/a\eta\sim \beta^{-1/3}$ is decreasing for a growing $\beta$ because of $a=1/H\sqrt2\sim(\tilde\beta \mathbb{N})^{1/2}$. However, as we will see below thermal corrections to CMB operationally depend on comoving scale parameters, and the high temperature limit corresponds to large $\beta$.

\section{Origin of inflation and formation of thermal CMB}
The gravitational instantons of the above type serve as a source of initial conditions for the cosmological evolution in the physical spacetime. Lorentzian signature Universes nucleate from the minimal surface of these instantons at $\tau_+$ --- the point of the maximal expansion of their Euclidean solutions. The latter when analytically continued to the complex plane by the rule $\tau=\tau_++it$ give the evolution in real Lorentzian time $t$. With $\dot a\equiv da/dt=ia'$ the generalized Friedmann equation (\ref{efeq}) when solved with respect to the Hubble factor takes the following form
    \begin{eqnarray}
    &&\frac{\dot a^2}{a^2}+\frac{1}{a^2}=
    \frac1B\left\{1-
    \sqrt{1-2B\left(\frac\varLambda3
    +\frac{C-B/2}{a^4}\right)}\right\}. \label{FriedmannL}
    \end{eqnarray}
Note that according to (\ref{bootstrap}), $C-B/2=(1/m_P^2)dF/d\eta$, the vacuum energy $B/2$ contained in $C$ does not contribute to the right hand side and, therefore does not gravitate \cite{bigboost}.

This equation admits the stage of quasi-exponential expansion driven by $\varLambda$, because the radiation very quickly dilutes in the course of expansion, $(C-B/2)/a^4\to 0$. This expansion can be interpreted as inflationary scenario consistent with observations provided it has appropriate exit from inflation with subsequent matter reheating and sufficiently high flatness of primordial perturbations spectra. This can be attained if we replace the constant $\varLambda$ by the potential of the {\em dynamical} slowly rolling inflaton field \cite{bigboost}, $\varLambda\to 8\pi GV(\phi)/3$, and add to (\ref{tree}) the kinetic term for $\phi$. Then inflation terminates by a conventional slow roll mechanism close to the minimum of the potential with $V(\phi)=0$, and reheating takes place due to inflaton oscillations in the vicinity of this minimum.

Addition of the dynamical field $\phi$ with a sufficiently flat potential replacing $\varLambda$ does not essentially change the microcanonical ensemble, because in the slow roll regime $\phi$ remains dynamically inert. In particular, in the Euclidean domain it also stays in the slow roll approximation, but in view of periodic boundary conditions it does not monotonically decrease. Rather, due to the oscillating scale factor and the friction force constantly changing sign, the inflaton follows these oscillations with a low amplitude and remains nearly constant during entire Euclidean evolution.

During inflationary stage a particle production of a conformally non-invariant matter takes over the polarization effects of conformal fields. After thermalization this matter gives rise to the energy density $\varepsilon$ which replaces in (\ref{FriedmannL}) the energy density of the primordial cosmological constant and primordial radiation, $\varLambda/3+(C-B/2)/a^4\to 8\pi G\varepsilon/3$. The resulting equation is essentially nonlinear in $\varepsilon$. But for sufficiently small matter density of newly born and thermalized particles, $8\pi BG\varepsilon/3\ll 1$, it satisfies the correspondence principle with Einstein GR. This guarantees a usual post-inflationary scenario.

Due to thermal nature of the initial quantum state all physical modes, including non-conformal inflaton and cosmological perturbations, have a spectrum modified by the factor
$1/(e^{k\eta}-1)$ (cf. Eq.(\ref{freeenergy}), we consider only bosonic fields). If we denote by $k$ the comoving momentum of the non-conformal modes -- contrary to the notation $\omega$ for conformal ones -- then the power spectrum of cosmological perturbations reads
    \begin{eqnarray}
    &&\delta_\phi^2(k)=
    \langle\, \hat\phi_k(t)
    \hat\phi_k(t)\,\rangle_{\rm thermal}\nonumber\\
    &&\qquad\qquad\qquad\qquad
    =
    \langle\, \hat a^\dagger_k
    \hat a_k+\hat a_k
        \hat a^\dagger_k\,\rangle_{\rm thermal}
    \,|u_k(t)|^2
    =|u_k(t)|^2\big(1+{2N_k(\eta)}\big),    \label{spectrum}
    \end{eqnarray}
where $u_k(t)$ is the positive frequency basis function in the $k$-mode and $N_k(\eta)$ is the occupation number in the thermal state with the ``comoving" temperature $1/\eta$
    \begin{eqnarray}
    &&N_k(\eta)=\frac1{e^{k\eta}-1}. \label{occupation}
	\end{eqnarray}
The corresponding spectral index acquires the thermal contribution originating from differentiating the thermal factor in (\ref{spectrum})
    \begin{eqnarray}
    &&n_s(k)=1+\frac{d}{d\ln k}\ln\delta_\phi^2(k)=n_s^{\rm vac}(k)+\Delta n_s^{\rm thermal}(k),\\
    &&\Delta n_s^{\rm thermal}(k)
    =\frac{d}{d\ln k}\ln\big(1+{2N_k(\eta)}\big).
    \end{eqnarray}
The vacuum part of $\delta_\phi^2(k)$ and $n_s(k)$ is of course determined by $|u_k(t)|^2$ which is nontrivial and essentially differs from the Einstein theory analogue, because the scalar and tensor sectors of CFT driven cosmology are more complicated than in the Einstein theory. They will be considered elsewhere, while the thermal contribution is universally determined by the $k$-dependence of the occupation number (\ref{occupation}) and derived below in terms of the observable wavenumber of CMB perturbations.

At present time, when the scale factor of the Universe equals $a_0$, the comoving wavenumber of the perturbation $k$ expresses in terms of its present physical wavenumber and wavelength
    \begin{eqnarray}
    k=a_0 k_{\rm phys}=\frac{a_0}{\lambda_{\rm phys}}
    \end{eqnarray}
Recalculating this quantity to the CMB multipole number $l$,  $\lambda_{\rm phys}(l)=\pi/l H_0$, we obtain in terms of the present horizon scale $H_0$ and the familiar cosmological density parameter $\Omega_0$
    \begin{eqnarray}
    k(l)\eta=l\,a_0 H_0\frac\eta\pi
    =\frac{l}{(\Omega_0-1)^{1/2}}\frac\eta\pi.
    \end{eqnarray}
This makes the Boltzmann factor $N_{k(l)}\simeq e^{-k(l)\eta}$ for observable CMB multipoles exponentially suppressed both by $l$ and a large value of the ratio of the horizon to curvature scale, $a_0 H_0=(\Omega_0-1)^{-1/2}\sim 10$. For a low temperature model with (\ref{cooleta}) it reads
    \begin{eqnarray}
    N_{k(l)}\simeq
    \left(\frac{3\beta}{16\pi^2\kappa^2}
    \right)^{l/\pi(\Omega_0-1)^{1/2}}\ll 1.                \end{eqnarray}
Even for a single-fold case with $\kappa=1$ this gives an absolutely negligible contribution both to $\delta^2_\phi(k)$ and $n_s(k)$. Within the low temperature limit this thermal contribution cannot be enhanced by increasing $\bar\beta$ -- a free parameter of the model, because this would bring the model beyond the low temperature approximation.

The effect of large $\tilde\beta$ is possible for high temperature limit with $\tilde\beta\gg 1$ suppressing the conformal time $\eta\sim 1/\tilde\beta^{1/6}$ according to (\ref{etahot}). In this limit the thermal contribution is weighted by the Boltzmann factor
    \begin{eqnarray}
    N_{k(l)}\simeq
    \exp\left[-\frac{2l}{(\Omega_0-1)^{1/2}}
    \left(\frac{\kappa^2}{180\,\tilde\beta}\right)^{1/6}\right]\simeq
    \exp\left[-
    \frac{10\,l}{(3\,\tilde\beta)^{1/6}}\,\kappa^{1/3}\right]               \end{eqnarray}
and can, in principle, be made $O(1)$ for a very large $\tilde\beta$. In particular, for $\kappa=1$ the thermal part of the CMB spectral index reads
    \begin{eqnarray}
    \Delta n_s^{\rm thermal}(k(l))\simeq
    2\,\frac{d\,N_k}{d\ln k}
    \simeq-2\,k\eta\,e^{-k\eta}\simeq
    -\frac{20\,l}{(3\,\tilde\beta)^{1/6}}e^{-10\,l/(3\,\tilde\beta)^{1/6}}.
    \end{eqnarray}
It is negative and, therefore, enhances the well known red tilt of the CMB spectrum generated by the vacuum contribution. Unfortunately, for low spins $s\leq 1$ the effect is still too small to be observable, because
    \begin{eqnarray}
    \frac1{60}\leq 3\,\tilde\beta_{\rm low-spin}\leq\frac{31}{30}\simeq 1,
    \end{eqnarray}
so that generation of the thermal correction in the third decimal order, $\Delta n_s^{\rm thermal}\sim -0.001$, would require $\tilde\beta\sim 10^6$.

Thus, only higher spin theories, $s\geq 3/2$, with much larger $\beta_s$ can qualitatively increase the thermal effect. The growth of $\beta_s$ with $s$ is a well known phenomenon \cite{ChristensenDuff} starting with lowest spins
    \begin{equation}
    \beta_s=\frac1{180}\times
    \left\{\begin{array}{cl}  1 &\quad s=0\\
     11 &\quad s=\frac12\\
     62 &\quad s=1
    \end{array}\right.                \label{betas}
    \end{equation}
and continuing for gravitino and gravitons which have $\beta_{3/2}=-235/180$ and $\beta_{2}=43/90$. Conformal non-invariant fields are, however, not suitable in our case, but Weyl invariant gravitino and graviton \cite{Tseytlinconf} maintain this tendency
    \begin{equation}
    \beta_s^{\rm Weyl}=\frac1{180}\times
    \left\{\begin{array}{cl} -548 &\quad s=\frac32\\
    1566 &\quad s=2\\
    {...} &\quad s>2\end{array}\right. \label{betas1}.
    \end{equation}
As it was recently advocated in \cite{Giombietal,Tseytlin13} these values of $\beta_s$ can be extended to higher spin conformal fields described by totally symmetric tensors and (Dirac) spin-tensors
    \begin{eqnarray}
    &&\beta_s=\frac1{360}\,\nu_s^2(3+14\nu_s),\quad \nu_s=s(s+1),\quad s=1,2,3,...\,,\label{boson}\\
    &&\beta_s=\frac1{720}\,
    \nu_s(12+45\nu_s+14\nu_s^2),\quad
    \nu_s=-2\Big(s+\frac12\Big)^2,\quad s=\frac12,\frac32,\frac52,...\,,   \label{fermion}
    \end{eqnarray}
where $\nu_s$ is their respective number of dynamical degrees of freedom (negative for fermions). Therefore, they contribute to the total value of $\tilde\beta$ specific values of $\beta_s$ per one degree of freedom, $\beta_s/|\nu_s|\sim s^4$, rapidly growing with spin.

These hypothetical conformal higher spin particles or their loop effects do not seem to be observable in the present Universe. Otherwise, enormously large value of $\beta\sim B$ in (\ref{FriedmannL}) would spoil the correspondence principle with GR in the present Universe. This seriously calls in question the possibility of the thermal mechanism with a large $\beta$. However, the formation of the thermal primordial spectrum took place in the early high-energy phase of field theory. If this phase is reachable by renormalization group flow of running coupling constants, then what can help the realization of this mechanism is the recently suggested {\em $a$-theorem} \cite{KomargodskiSchwimmer,Komargodski}.

\section{The $a$-theorem and dilaton mode in CFT cosmology}
The possibility of climbing up the ladder of higher spins, and thus increasing the coefficient $\beta$ in the CFT model, can be associated with the renormalization group effects in interacting conformal field theory. As has recently been persuasively advocated on the basis of the trace anomaly matching  \cite{matching}, the renormalization group flow from ultraviolet (UV) to infrared (IR) limit in four dimensions is subject to the so-called ``$a$-theorem" \cite{KomargodskiSchwimmer,Komargodski}. This theorem is the 4D analogue of the two-dimensional $c$-theorem of Zamolodchikov \cite{Zamolodchikov}. It represents the statement of the decreasing Gauss-Bonnet coefficient in the trace anomaly (\ref{anomaly}) in the course of this flow from UV to IR. In notations of \cite{KomargodskiSchwimmer,Komargodski}, using the Lorentzian signature spacetime and the corresponding Lorentzian effective action $W$ (see Appendix A), the trace anomaly (\ref{anomaly}) reads
    \begin{eqnarray}
    &&\langle\, T^\mu_\mu\,\rangle\equiv
    \frac2{g^{1/2}}\,g^{\mu\nu}\frac{\delta
    W}{\delta g^{\mu\nu}} =
    \mbox{\boldmath$a$} E -\mbox{\boldmath$c$}\, C_{\mu\nu\alpha\beta}^2
    +\mbox{\boldmath$b$}\, \Box R.         \label{Wanomaly}
    \end{eqnarray}
According to \cite{KomargodskiSchwimmer,Komargodski} the difference between the UV and IR values of this parameter is related to the total cross section $\sigma(s)=s\,{\rm Im}\, {\cal A}(s,t)_{t=0}>0$ of the forward scattering of the dilaton -- Nambu-Goldstone boson of broken conformal symmetry,
    \begin{eqnarray}
    \mbox{\boldmath$a$}_{UV}-\mbox{\boldmath$a$}_{IR}=
    \frac1{4\pi}\int_{s>0} ds\,
    \frac{\sigma(s)}{s^2}.        \label{dispersionrelation}
    \end{eqnarray}
The positivity of $\sigma(s)$ in unitary theory guarantees here the positive increment $\mbox{\boldmath$a$}_{UV}-\mbox{\boldmath$a$}_{IR}>0$.

Important point of this statement is the fact that the Gauss-Bonnet invariant itself, being a total derivative part of UV divergences of the theory, never contributes to its local dynamics and seemingly does not lead to any positivity restrictions. Indeed, in contrast to $E$-invariant, the Weyl squared part of the effective action contributes not only to UV divergences (say, in dimensional regularization with $d\to 4$), which are just the integrated conformal anomaly (\ref{Wanomaly}), but also to their finite tail -- the logarithmic nonlocal part (cf. Appendix A, $d^4x_L=dx^0 d^3{\bf x}$ -- Lorentzian spacetime integration measure)
    \begin{eqnarray}
    &&iW=\frac{i}{2(4\pi)^2}\frac1{2-\frac{d}2}\int d^4x_L\,g^{1/2}\Big(\mbox{\boldmath$c$}\, C_{\mu\nu\alpha\beta}^2-\mbox{\boldmath$a$} E\Big)\nonumber\\
    &&\qquad\qquad\qquad
    -\frac{i}{2(4\pi)^2}\int d^4x_L\,g^{1/2}\left(\mbox{\boldmath$c$}\, C_{\mu\nu\alpha\beta}
    \ln\frac{-\Box
    -i\varepsilon}{\mu^2}C^{\mu\nu\alpha\beta}
    +...\right).                        \label{W}
    \end{eqnarray}
In the momentum representation the nonlocal logarithm acquires the local imaginary part,
    \begin{eqnarray}
    \ln\frac{-\Box
    -i\varepsilon}{\mu^2}=
    \ln\frac{|p^2|}{\mu^2}-i\pi\theta(-p^2),
    \end{eqnarray}
so that the imaginary part of the effective action takes the form quadratic in Fourier transform of Weyl tensor $\hat C_{\mu\nu\alpha\beta}(p)$
    \begin{eqnarray}
    {\rm Im}\, W=\frac1{32\pi}\int d^4p\,\Big(\mbox{\boldmath$c$}\; |\hat C_{\mu\nu\alpha\beta}(p)|^2\,\theta(-p^2)+...\Big).
    \end{eqnarray}
Then, unitarity of the theory, $|\exp(iW)|<1$ demanding ${\rm Im} W>0$, immediately leads to positive definiteness of the coefficient $\mbox{\boldmath$c$}$. No such a restriction holds for the coefficient of $E$, because the Gauss-Bonnet term in the divergent part of the action (\ref{W}) in view of its total derivative nature does not have a logarithmic counterpart among finite nonlocal terms of $W$. Nevertheless, the Gauss-Bonnet coefficient $\mbox{\boldmath$a$}$ is not dynamically inert but rather effects the scattering of the dilaton field -- the parameter of broken local Weyl invariance. As shown in \cite{KomargodskiSchwimmer,Komargodski}, unitarity of this scattering process gives rise to the restriction on the RG flow of $\mbox{\boldmath$a$}$ which is more complicated than a simple restriction on the sign of Weyl coefficient $\mbox{\boldmath$c$}$. This dilaton field and its action induced by $E$-part of the trace anomaly (\ref{anomaly}) both arise as a consequence of this symmetry breakdown and can be derived by the Wess-Zumino procedure of anomaly integration.

Similar situation occurs in our CFT driven cosmology. Here the coefficient of the topological term $B=3\beta/4 m_P^2$ is crucially important both for the dynamics of the cosmological background (via (\ref{efeq})) and the CMB power spectrum. However, in contrast to the usual CFT setup using the spacetime metric merely as an auxiliary tool which probes the correlators of  the CFT stress tensor,  here the dilaton field is a well-known physical observable -- the cosmological scale factor. Correspondingly, the logic of the $a$-theorem application in cosmology follows from the fact that the cosmological expansion can be associated with the transition from deep UV to IR regimes.\footnote{Which of course should not be interpreted literally by directly relating the decreasing RG running scale $\mu$ in (\ref{dispersionrelation}) to the growing scale factor $a$ (see discussion in \cite{WoodardRG}), but rather understood as effective action describing early and late cosmology respectively with the values $\mbox{\boldmath$a$}_{UV}$ and $\mbox{\boldmath$a$}_{IR}$ of the $\mbox{\boldmath$a$}$-parameter. Concrete realization of this RG flow still remains to be done similarly to other examples of cosmological RG implications like \cite{RG}, where $\mu$ was related to the inflaton field value.} Moderately small value of $a$ in late cosmology can be a result of the evolution from a much larger initial value $a_{UV}$ responsible for the formation of a considerable thermal part of CMB. This obviously follows from the $a$-theorem trace anomaly (\ref{Wanomaly}) whose coefficients $\mbox{\boldmath$a$}$, $\mbox{\boldmath$b$}$ and $\mbox{\boldmath$c$}$ are related to the coefficients of (\ref{anomaly})
    \begin{eqnarray}
    \mbox{\boldmath$a$}=\frac\beta{32\pi^2},\quad
    \mbox{\boldmath$b$}=\frac\alpha{32\pi^2},\quad
    \mbox{\boldmath$c$}=-\frac\gamma{32\pi^2},
    \end{eqnarray}
so that large $\mbox{\boldmath$a$}_{UV}$ implies large $\beta$ in early cosmology at its nucleation from the cosmological instanton.

Dynamics of the dilaton field $\sigma$, whose scattering cross section guarantees the positivity of $\beta_{UV}-\beta_{IR}=32\pi^2 (\mbox{\boldmath$a$}_{UV}-\mbox{\boldmath$a$}_{IR})$, is governed by the action which can be obtained by the Wess-Zumino procedure of integrating the trace anomaly (\ref{anomaly}) along the orbit of the conformal group
    \begin{eqnarray}
    &&g_{\mu\nu}=e^\sigma \bar g_{\mu\nu},   \label{orbit}\\
    &&\frac{\delta \varGamma[\,e^\sigma\bar g\,]}{\delta\sigma}\,=\frac{1}{4(4\pi)^2}\,g^{1/2}
    \big(\alpha \Box R +
    \beta E +\gamma C_{\mu\nu\alpha\beta}^2\big)
    \Big|_{\;g\,=\,e^\sigma\bar g}\,.        \label{orbiteq}
    \end{eqnarray}
Here we reproduce this integration in the Euclidean version of the theory.\footnote{2D $c$-theorem is associated with the reflection positivity in Euclidean QFT, whereas the 4D version of this theorem is better interpreted from the viewpoint of unitarity of the theory in the Lorentzian spacetime \cite{Komargodski}. Thus, the Komargodski-Schwimmer $a$-theorem (\ref{Wanomaly})-(\ref{dispersionrelation}) is formulated in the Lorentzian spacetime with the effective action $W$ related to the Euclidean effective action $\varGamma$ by Wick rotation $iW=-\varGamma$. In contrast to the Euclidean CFT, dilaton scattering and its cross section are defined in the Lorentzian spacetime and incorporate a familiar notion of unitarity. In comparison with a formal CFT where the Euclidean formulation is used merely as a calculational trick, here we have the analytic junction of the Euclidean theory on the cosmological instanton with the theory of expanding universe in Lorentzian spacetime. This naturally unifies Euclidean and Lorentzian spacetime versions of one theory into a single entity.} The resulting Wess-Zumino action for $\sigma$ is just the difference of effective actions calculated on two members of this orbit $g_{\mu\nu}$ and $\bar g_{\mu\nu}$. It reads \cite{anomalyaction,BMZ}
    \begin{eqnarray}
    &&\varGamma[\,g\,]-\varGamma[\,\bar g\,]=
    \frac{1}{2(4\pi)^2}\int d^4x \bar g^{1/2} \left\{\,\frac{1}{2}\,
    \Big[\,\gamma\, \bar C_{\mu\nu\alpha\beta}^2
    +\beta\,\Big(\bar E-\frac{2}{3}\,\bar\Box \bar R\Big)\Big]\,
    \sigma
    +\,\frac{\beta}{2}\,\sigma{\cal \bar D}\sigma\,\right\}\nonumber\\
    &&\qquad\qquad\qquad\qquad-
    \,\frac{1}{2(4\pi)^2}
    \Big(\frac{\alpha}{12}
    +\frac{\beta}{18}\Big)\,
    \int d^4x\,\Big(g^{1/2}R^2(g)-
    \bar{g}^{1/2}R^2(\bar{g})\Big),              \label{RTF}
    \end{eqnarray}
where all barred quantities are built in terms of $\bar g_{\mu\nu}$ and ${\cal D}$ is the fourth-order operator
    \begin{eqnarray}
    {\cal D} = \Box^2 + 2R^{\mu\nu}\nabla_{\mu}\nabla_{\nu} -
    \frac{2}{3} R\,\Box + \frac{1}{3}(\nabla^{\mu}R)\,\nabla_{\mu}.   \label{}
    \end{eqnarray}
This operator has a number of special properties, including the local Weyl invariance of its densitized version
    $\bar g^{1/2}{\cal\bar D}=g^{1/2}{\cal D}$
and the linear conformal transformation law for the Gauss-Bonnet density (modified by the $\Box R$ term)
    \begin{eqnarray}
    g^{1/2}\Big(E-\frac23\,\Box R\,\Big)=
    \bar g^{1/2}\Big(\bar E-\frac23\,\bar\Box\bar R\,\Big)
    +2\,\bar g^{1/2}{\cal\bar D}\sigma,\,\,\,\,\,
    g_{\mu\nu}=e^\sigma \bar g_{\mu\nu}                     \label{basic}
    \end{eqnarray}
The absence of local Weyl invariance of this quantity -- ``non-abelian" nature of the Gauss-Bonnet anomaly -- is a main source of the nontrivial Wess-Zumino action \cite{KomargodskiSchwimmer}.

It is important to notice that functional integration of Eq.(\ref{orbiteq}) a priori leads to the answer which has the form of a fourth-order polynomial in $\sigma$. Remarkable property of this polynomial is, however, that all cubic and quartic terms can be collected into the curvature squared invariant \cite{BMZ}, $\int d^4x\,\big(g^{1/2}R^2(g)-\bar g^{1/2} R^2(\bar g)\big)$, which forms the second line of (\ref{RTF}). This in turn implies that after a finite renormalization of the action by this local counterterm,
    \begin{eqnarray}
    \varGamma[\,g\,]\to \varGamma_{R}[\,g\,]
    =\varGamma[\,g\,]
    +\frac1{2\,(4\pi)^2}\,
    \frac\alpha{12}\int d^4x\,
    g^{1/2}\,R^2(g),                   \label{renormaction}
    \end{eqnarray}
the increment of this action along the orbit of the local conformal group (\ref{orbit}) becomes $\alpha$-independent and acquires the following {\em minimal} form
    \begin{eqnarray}
    &&\varGamma_R[\,g\,]-\varGamma_R[\,\bar g\,]=
    \frac{\gamma}{4(4\pi)^2}\int d^4x \bar g^{1/2}\,\sigma
    \,\bar C_{\mu\nu\alpha\beta}^2             \nonumber\\
    &&\qquad\qquad\qquad\qquad
    +\frac{\beta}{2(4\pi)^2}\int d^4x \bar g^{1/2} \left\{\,\frac{1}{2}\,
    \sigma \bar E-\Big(\bar R^{\mu\nu}-\frac12\bar g^{\mu\nu}\bar
    R\,\Big)\,\partial_\mu\sigma\,\partial_\nu\sigma \right.  \nonumber\\
    &&\qquad\qquad\qquad\qquad\qquad\qquad\qquad\qquad\left.-
    \,\frac12\,\bar\Box\sigma\,
    (\bar\nabla^\mu\sigma\,\bar\nabla_\mu\sigma)
    -\frac18\,(\bar\nabla^\mu\sigma\,
    \bar\nabla_\mu\sigma)^2\right\},         \label{minimal}
    \end{eqnarray}
where again all barred quantities are built in terms of the metric $\bar g_{\mu\nu}$. This property is, of course, equivalent to the well-known statement that the coefficient of $\Box R$ in the trace anomaly can always be renormalized to zero by the counterterm quadratic in Ricci scalar \cite{BD}, which is admissible from the viewpoint of UV renormalization due to its locality.

The renormalization (\ref{renormaction}) has another important consequence -- with $\alpha=0$ the terms with higher order (quartic) derivatives of $\sigma$, contained in the combination $\sigma\bar{\cal D}\sigma-\frac19e^{2\sigma}R^2(e^\sigma\bar g)$, completely cancel out, and the resulting {\em minimal} Wess-Zumino action (\ref{minimal}) does not acquire extra hihger-derivative degrees of freedom \cite{slih}. This might not be obvious for the third term in curly brackets of Eq.(\ref{minimal}), but its variation shows that it contributes to equations of motion maximum second order derivatives of $\sigma$
    \begin{eqnarray}
    \frac{\delta}{\delta\sigma}
    \int d^4x \bar g^{1/2}\,\bar\Box\sigma\,
    (\bar\nabla^\mu\sigma\,\bar\nabla_\mu\sigma)=
    2\,\bar g^{1/2}\Big((\bar\nabla_\mu\bar
    \nabla_\nu\sigma)^2
    -(\bar\Box\sigma)^2
    +\bar R^{\mu\nu}\bar\nabla_\mu
    \sigma\bar\nabla_\nu\sigma\Big)    \label{sigmaeq}
    \end{eqnarray}
and supplies the inverse propagator of the dilaton with the terms which in the presence of a nontrivial inhomogeneous background, $\nabla_\mu\sigma\neq 0$, force its characteristic surface (sound cone) to deviate from the light cone,
    \begin{eqnarray}
    &&\frac{\delta^2}{\delta\sigma(y)\,\delta\sigma(x)}
    \int d^4x \bar g^{1/2}\,\bar\Box\sigma\,
    (\bar\nabla^\mu\sigma\,\bar\nabla_\mu\sigma)\nonumber\\
    &&\qquad\qquad\qquad=
    4\,\bar g^{1/2}\Big[\big(\bar\nabla^\mu\bar\nabla^\nu\sigma-\bar g^{\mu\nu}\bar\Box\sigma\big)
    \bar\nabla_\mu\bar\nabla_\nu+\bar R^{\mu\nu}\bar\nabla_\mu\sigma\,
    \bar\nabla_\nu\Big]\, \delta(x,y).   \label{kinetic}
    \end{eqnarray}
This cubic term plays a special role in recent modifications of gravity theory \cite{DGP,boxnablapi} like brane induced gravity models and massive graviton models where it survives the so-called decoupling limit and represents the ghost-free higher-derivative braiding of metric and matter \cite{Vikmanetal}.\footnote{Though the kinetic term (\ref{sigmaeq}) of the variational equation for $\sigma$ is quadratic in $\nabla\nabla\sigma$, $\ddot\sigma$ enters it linearly which might be important for the Cauchy problem of this field and its dynamical nature discussed below, cf. footnote \ref{footnote}.}

To the best of our knowledge, the minimal version of the dilaton action in the form (\ref{minimal}) was first discussed in \cite{matching}. Then it was used in the derivation of the $a$-theorem in \cite{KomargodskiSchwimmer}, unitarity (and, therefore, scattering cross section positivity, $\sigma(s)>0$, in (\ref{dispersionrelation})) of the dilaton contribution being guaranteed by the absence of higher-derivative ghosts in (\ref{minimal}).\footnote{The term of (\ref{minimal}) linear in the Einstein tensor was also used to probe the conical singularity in spacetime associated with the entanglement entropy and the dilaton contribution to the latter \cite{Solodukhin}.} The dynamical nature of the dilaton and its actual contribution to $\mbox{\boldmath$a$}_{IR}$ was, however, retracted in the later paper \cite{Komargodski} where it was assumed to be merely an external field never forming quantum loops. In our cosmological context the dilaton $\sigma$ acquires the meaning of a real physical observable -- the logarithm of the cosmological scale factor $a$ in the FRW metric\footnote{We use this notation for the cosmological scale factor, which is very close to the boldfaced notation for the coefficient of the Gauss-Bonnet anomaly in (\ref{Wanomaly}), but hope that this will not lead to a confusion. The same concerns overlap of notations for $\sigma(s)$ -- the dilaton scattering cross section -- and the dilaton field itself $\sigma=\sigma(\tau)$, which can easily be distinguished by context.} and the conformal mode of the metric perturbations. Therefore, one might expect a dynamical input from quantum and thermal dilaton fluctuations. However, in Einstein theory the dilaton (or conformal) mode is nondynamical because it gets eliminated by the Hamiltonian constraint of the theory. The same mechanism is likely to hold here, though perhaps by the price of essential nonlinearity and non-analyticity in the resulting constraints\footnote{\label{footnote}Unlike in Einstein theory, the variation of the anomaly action with respect to the lapse function $N$ contains the second order time derivative $\ddot\sigma$. In order to convert this equation into the constraint reducing the number of degrees of freedom, $\ddot\sigma$ has to be expressed in terms of $\sigma$ and its spatial derivatives from the variational equation for $\sigma$. Though the latter is linear in $\ddot\sigma$ (cf. Eq.(\ref{sigmaeq})), the inversion of its coefficient proportional to curvature and spatial gradients of $\sigma$ brings very nonlinear and non-analytic structures.}. We will demonstrate this mechanism in the long wavelengths limit of the scalar sector of cosmological variables. It corresponds to the minisuperspace approximation of the FRW metric (\ref{FRW}) and spatially homogeneous dilaton $\sigma(\tau)$.

In this case $\bar g_{\mu\nu}$ should be identified with the metric of the Einstein static Universe of a unit radius, $d\bar s^2=d\eta^2+d\Omega_3^2$, where $\eta$ is a conformal time $\eta=\int d\tau\,N/a$ (cf. (\ref{period})) and the dilaton conformal factor relating the two metrics (\ref{orbit}) is the scale factor of (\ref{FRW}), $e^\sigma=a^2$. Simple calculation shows that the anomaly part of the action (\ref{minimal}) when expressed in terms of the original FRW variables does not contain second-order derivatives of $a$,
    \begin{eqnarray}
    &&\varGamma_R[\,g\,]-\varGamma_R[\,\bar g\,]=\frac{3\beta}4\int_{S^1} d\tau\,N \left(\frac{a'^2}{a}
    -\frac{a'^4}{6 a}\right),         \label{anomalyaction}
    \end{eqnarray}
because the cubic term $\bar\Box\sigma(\bar\nabla\sigma)^2\sim \ddot\sigma\dot\sigma^2$ reduces to the total derivative.
The calculation of $\varGamma_R[\,\bar g\,]$ confirms the same property, because it yields a typical boson or fermion statistical sum (\ref{freeenergy}) of a free field theory on a static $S^1\times S^3$ spacetime of the Euclidean time period (\ref{period}) and unit radius plus the contribution of the vacuum Casimir energy $E_{\rm vac}$ \cite{slih}.

Minimal form of the anomalous dilaton action with the coefficient $\alpha$ renormalized to zero yields yet another interesting property -- a particular value of this Casimir energy. Namely, this value turns out to be universally expressed in terms of a single Gauss-Bonnet anomaly coefficient $\beta$ \cite{slih}. It is well known that for lowest CFT spins this Casimir energy with covariantly renormalized UV infinities expresses in terms of {\em two} trace anomaly coefficients $\alpha$ and $\beta$ \cite{universality}
    \begin{eqnarray}
    E_{\rm vac} =\sum\limits_\omega\frac\omega2\,\Big|_{\;\rm renorm}=\frac{3\beta-\alpha/2}8.
    \end{eqnarray}
But in addition we had to perform a finite renormalization (\ref{renormaction}) by a curvature squared counterterm putting $\alpha$ to zero and thus eliminating extra higher-derivative dilaton mode. This means that the total renormalized effective action on the static Einstein universe acquires the extra term,
    \begin{eqnarray}
    &&\varGamma_R[\,\bar g\,]=F(\eta)+\eta\, E_{\rm vac}+\frac1{2\,(4\pi)^2}\,
    \frac\alpha{12}\int d^4x\,
    \bar g^{1/2}\,R^2(\bar g)=
    F(\eta)+\frac{3\beta}8\eta.               \label{renorm1}
    \end{eqnarray}
This automatically renders the Casimir energy $\alpha$-independent and contributes a term $m_P^2B/2a$ to the effective action (\ref{effaction}).

Taken together (\ref{anomalyaction}) and (\ref{renorm1}) contribute to the total effective action of the CFT cosmology (\ref{effaction}) which generates the Hamiltonian constraint (\ref{efeq}), thus indicating the absence of dynamical modes in the minisuperspace and the long wavelengths scalar sector of CMB.

The logic of the $a$-theorem application is that the cosmological expansion can be associated with the transition from deep UV to IR regimes. The RG running in interacting and gravitating CFT can lead to the redistribution in the UV limit of the full set of conformal degrees of freedom to a higher spin domain, possessing according to the $a$-theorem higher values of $\beta$ and $\tilde\beta$.  The present value of $\beta_{IR}=32\pi^2\mbox{\boldmath$a$}_{IR}$ can be a result of the evolution from a much larger initial value $\beta_{UV}$ responsible for the formation of a considerable thermal part of CMB.

\section{Conclusions}

The last WMAP and new Planck data on the CMB power spectrum and its non-gaussianities \cite{WMAP9} justify interest in variety of modified vacuum and non-vacuum states of cosmological perturbations \cite{nongaussianity,GancKomatsu,Ashoorioonetal}. In this context the microcanonical state in the CFT cosmology has an advantage that it comes from first principles of quantum gravity \cite{slih,why} rather than from some ad hoc assumptions. Quite remarkably its formalism and physical predictions are determined by the Gauss-Bonnet anomaly and tightly related to the dilaton dynamics associated with the $a$-theorem -- a higher-dimensional extension of classical $c$-theorem.  In particular, stronger ``heating" of the CMB spectrum can be mediated by the RG flow interpolating between the UV and IR stages of cosmological expansion and, in view of this theorem, shifting the CFT model in UV to a higher spin phase. This opens a number of prospects for a further research.

A remarkable convolution of properties -- minimal form of the anomalous dilaton action, elimination of higher-derivative dilaton modes, non-dynamical nature of the dilaton (or scalar) sector of CMB as a consequence of turning on dynamical gravity and, finally, a particular value of the Casimir energy -- is not yet fully understood in context of the $a$-theorem and its cosmological applications. In this theorem the dilaton plays an auxiliary role of a fictitious external field, whereas in gravity and cosmology it is a physical observable which is a part of gravitational equations of motion and cannot be disregarded by hands.

On the other hand, the dilaton action invokes serious issues of causality and locality. As advocated in \cite{superluminality} the structure of the dilaton kinetic terms (\ref{kinetic}) can lead to superluminal propagation of its modes, and no consistent UV completion of the theory is possible unless certain positivity bounds are satisfied by the coefficients of the low-energy effective action. One such bound -- negative value of the coefficient of $(\bar\nabla^\mu\sigma\,\bar\nabla_\mu\sigma)^2$ in (\ref{minimal}) (positivity in the Lorentzian effective action of \cite{superluminality}) -- is satisfied, but a more convincing argument in favor of our CFT driven cosmology can be a complete elimination of the dilaton from the sector of propagating modes. This argument seems working here, because turning on dynamical gravity is likely to do this via the Hamiltonian constraint at least in the long wavelengths limit.
All this is important for the CMB non-gaussianity, because the vacuum part determined by $|u_k(t)|^2$ in (\ref{occupation}) reveals speed of sound phenomena \cite{superluminality} since the inflaton dynamics is strongly mediated by $\sigma$ with its nontrivial sound cone in (\ref{kinetic}).

Here we dwelled on the thermal input into the power spectrum of primordial perturbations. For all low spin CFT models it turned out to be exponentially suppressed due to large curvature scale, $\sim(\Omega_0-1)^{-1/2}\ll 1$, and is beyond current observations. However, hypothetical conformal models of higher spin interactions, which seem to be inevitable in a unified picture of the early quantum Universe, can lead to the enhancement of this thermal effect. The key to this phenomenon is the mechanism of the $a$-theorem for the RG flow between the early UV phase of the Universe and its present IR regime. The efficiency of this mechanism should be tested, as its RG increment $\mbox{\boldmath$a$}_{UV}-\mbox{\boldmath$a$}_{IR}$ is advocated to be always bounded \cite{PolchinskiRattazzi}.

Moreover, {\em interacting} higher-spin conformal fields do not seem to be explicitly known yet, except conformal gravitino with $s=3/2$ and Weyl graviton with $s=2$. Recent progress in generalizing these  models to conformal higher spin fields of arbitrary $s$ on the Einstein-space background allowed one to compute their 1-loop Weyl anomaly coefficients (\ref{boson})-(\ref{fermion}) (by indirect AdS/CFT method in \cite{Giombietal} and directly in \cite{Tseytlin13}). But this result still leaves the issue of unitarity violation caused by inevitable higher derivatives in wave operators of these fields (cf. $\beta_{3/2}<0$ in (\ref{betas}) for third order operator and negative values of $\beta_s$ in (\ref{fermion}) for fermions with $\nu_s<0$ and $2s$ derivatives in the wave operator \cite{Tseytlinconf}). Thus, the progress here strongly depends on advancing theory of conformal higher spin models \cite{Tseytlinconf,higherspin,Maldacena,Giombietal,Tseytlin13}. There is a lot more to be learned within this remarkable interplay between fundamental conformal invariance, the $a$-theorem and physics of the very early Universe.

\appendix
\renewcommand{\thesection}{Appendix \Alph{section}.}
\renewcommand{\theequation}{\Alph{section}.\arabic{equation}}

\section{Conformal anomaly conventions}

Signs of the conformal anomaly coefficients are very important for the dynamics of the CFT cosmology. On the other hand, various works treating conformal anomalies \cite{Duffanomaly,BD,KomargodskiSchwimmer} operate with different sign conventions in the definition of metric stress tensor, metric signature and Lorentzian vs Euclidean signatures of spacetime. Here we present a brief overview of basic sign conventions and relations for local Weyl anomaly in low spin theories.

A conventional relation between the Lorentzian effective action $W$ and its Euclidean counterpart $\varGamma$ under the Wick rotation is
    \begin{eqnarray}
    iW=-\varGamma,
    \end{eqnarray}
whence we have for their metric variations
    \begin{eqnarray}
    i\int d^4x_L\,\frac{\delta W}{\delta g_{\mu\nu}}\,\delta g_{\mu\nu}=-\int d^4x_E\,\frac{\delta\varGamma}{\delta g_{\mu\nu}}\,\delta g_{\mu\nu},
    \end{eqnarray}
where the Lorentzian and Euclidean spacetime integration measures are related by $d^4x_L=-i\,d^4x_E$. Therefore, with the conventional definition for the Lorentzian theory stress tensor $T_{\mu\nu}^L$ \cite{BD,KomargodskiSchwimmer,Komargodski} and that of the Euclidean theory $T^{\mu\nu}_E$ \cite{Tseytlinconf},
    \begin{eqnarray}
    &&\langle\, T_{\mu\nu}^L\rangle\equiv
    \frac2{g^{1/2}}\frac{\delta
    W}{\delta g^{\mu\nu}},\\
    &&\langle\, T^{\mu\nu}_E\rangle\equiv
    \frac2{g^{1/2}}\frac{\delta
    \varGamma}{\delta g_{\mu\nu}}
    \end{eqnarray}
(note the difference in the position of indices), we have a formal equality of Lorentzian and Euclidean trace anomalies as local functions of their respective metrics
    \begin{eqnarray}
    \langle\, T^\mu_{\,\mu\,E}\,\rangle=\langle\, T^\mu_{\,\mu\,L}\,\rangle.
    \end{eqnarray}
This relation is of course independent of the signature choice $(-+++)$ or $(+---)$ in the Lorentzian case.

In conformally invariant theories the UV divergences of the one-loop effective action and the trace anomaly are both given by $a_2(x)$ --- the trace over isotopic indices of the coincidence limit of the second Schwinger-DeWitt coefficient \cite{SchDW} (or the fourth Gilkey-Seely coefficient $b_4(x)$)
    \begin{eqnarray}
    &&\varGamma^{\rm div}=
    -\frac1{(4\pi)^2}\,
    \frac1{4-d}\int d^4x\,g^{1/2}a_2(x),   \label{div}\\
    &&\langle\, T^\mu_\mu\,\rangle\equiv
    \frac2{g^{1/2}}g_{\mu\nu}\frac{\delta
    \varGamma}{\delta g_{\mu\nu}}=-\frac1{(4\pi)^2}\,a_2(x),
    \end{eqnarray}
where $d\to 4$ is the parameter of the dimensional regularization.

For conformally invariant fields of lowest spins $a_2(x)$ reads
    \begin{eqnarray}
    \frac1{(4\pi)^2}\,a_2=
    \mbox{\boldmath$c$}\, C_{\mu\nu\alpha\beta}^2-\mbox{\boldmath$a$} E
    -\mbox{\boldmath$b$}\, \Box R,
    \end{eqnarray}
where the coefficients are contributed by $\mathbb{N}_0$ real scalars, $\mathbb{N}_{1/2}$ Dirac spinors and $\mathbb{N}_{1}$ vector multiplets (including relevant contributions of Faddeev-Popov ghosts subtracting temporal and longitudinal polarizations)
    \begin{eqnarray}
    &&\mbox{\boldmath$a$}=\frac1{360\,(4\pi)^2}\, \big(\mathbb{N}_0+11 \mathbb{N}_{1/2}+
    62 \mathbb{N}_{1}\big),                \label{101}\\
    &&\mbox{\boldmath$c$}=\frac1{120\, (4\pi)^2}\, \big(\mathbb{N}_0+6\mathbb{N}_{1/2}+
    12 \mathbb{N}_{1}\big),\\
    &&\mbox{\boldmath$b$}=-\frac1{180\, (4\pi)^2}\, \big(\mathbb{N}_0+6\mathbb{N}_{1/2}+
    12 \mathbb{N}_{1}\big).
    \end{eqnarray}
In the dimensional regularization the coefficient of $\Box R$, $\mbox{\boldmath$b$}$, is related to $\mbox{\boldmath$c$}$ by the equation $\mbox{\boldmath$b$}=-\frac23\,\mbox{\boldmath$c$}$, but in the  zeta-function regularization this relation does not hold for a vector multiplet, $s=1$, and should be replaced by $\mbox{\boldmath$b$}_1=-\mbox{\boldmath$c$}_1$.

In the approximation quadratic in spacetime curvature the finite nonlocal part of the Euclidean effective action can be obtained from (\ref{div}) by replacing the divergent factor $1/(4-d)$ with the nonlocal operator \cite{CPT}
    \begin{eqnarray}
    F(\Box)=\frac1{4-d}\,
    \left(\frac{-\Box}{\;\,\mu^2}\right)^{\frac{d}2-2}
    =\frac1{4-d}-\frac12\ln\frac{-\Box}{\;\,\mu^2},
    \end{eqnarray}
which in the Lorentzian theory leads to (\ref{W}) on account of the Wick rotation.

\section*{Acknowledgements}
I am grateful to J.Garriga and S.Sibiryakov for helpful discussions and acknowledge support at the workshop YITP-T-12-03 ``Gravity and Cosmology 2012" of the Yukawa Institute for Theoretical Physics. This work was supported by the RFBR grant No. 11-02-00512.

\end{document}